\pdfoutput=1

\newif\ifarxiv
\ifdefined\duneuronoarxiv
\arxivfalse
\else
\arxivtrue
\fi

\arxivtrue

\documentclass[10pt]{article}

\usepackage{amsmath,amssymb}

\usepackage{changepage}

\usepackage[utf8x]{inputenc}

\usepackage{textcomp,marvosym}

\usepackage{cite}

\usepackage{nameref,hyperref}

\usepackage[right]{lineno}

\usepackage{microtype}
\DisableLigatures[f]{encoding = *, family = * }

\usepackage[table]{xcolor}

\usepackage{array}

\usepackage{multirow}

\newcolumntype{+}{!{\vrule width 2pt}}

\usepackage[labelformat=simple]{subcaption}
\usepackage{siunitx}
\usepackage{float}
\usepackage{listings}

\usepackage{stmaryrd}
\usepackage{relsize}
\usepackage{physics}
\usepackage{tikz}
\usetikzlibrary{positioning,shapes,shadows,arrows}

\newlength\savedwidth

\usepackage{todonotes}
\reversemarginpar

\usepackage[aboveskip=5pt,labelfont=bf,labelsep=period,justification=raggedright,singlelinecheck=off]{caption}

\makeatletter
\renewcommand{\@biblabel}[1]{\quad#1.}
\makeatother

\newcommand{\scp}[2]{\left\langle #1 , #2 \right\rangle}

\newcommand\Cpp{C\nolinebreak[4]\hspace{-.05em}\raisebox{.4ex}{\relsize{-3}{\textbf{++}}}\xspace}
\newcommand{\Oo}{\mathcal{O}}
\newcommand{\Tt}{\mathcal{T}}
\newcommand{\NN}{\mathbb{N}}
\newcommand{\RR}{\mathbb{R}}
\newcommand{\dipoleposition}{x_{\text{dp}}}
\newcommand{\dipolemoment}{M}
\newfloat{lstfloat}{t}{lop}
\floatname{lstfloat}{Listing}
\lstdefinestyle{pythoncode}{
  language=Python,
}
\lstdefinestyle{matlabcode}{
  breaklines=true,
  identifierstyle=\color{black},
  stringstyle=\color{black},
  commentstyle=\color{gray},
  showstringspaces=false,
  emph=[1]{for,end,break},
  emphstyle=[1]\color{black}
}

\usepackage{blindtext}
\usepackage{titling}

\title{DUNEuro - A software toolbox for forward modeling in bioelectromagnetism}

\begin{document}

\vspace*{0.2in}

\begin{flushleft}
{\Large
\textbf\newline{\thetitle}
}
\newline
\\
Sophie Schrader\textsuperscript{1\YinYang},
Andreas Westhoff\textsuperscript{1,2,3,\textcurrency\YinYang},
Maria Carla Piastra\textsuperscript{1,2,4},
Tuuli Miinalainen\textsuperscript{5,6},
Sampsa Pursiainen\textsuperscript{5},
Johannes Vorwerk\textsuperscript{1,7},
Heinrich Brinck\textsuperscript{3},
Carsten H. Wolters\textsuperscript{1,8},
Christian Engwer\textsuperscript{2*}
\\
\bigskip
\textbf{1} Institute for Biomagnetism and Biosignalanalysis, University of M\"unster, Germany
\\
\textbf{2} Applied Mathematics: Institute for Analysis and Numerics, University of M\"unster, Germany
\\
\textbf{3} Westf\"alische Hochschule, University of Applied Sciences Recklinghausen, Germany
\\
\textbf{4} Radboud University Nijmegen Medical Centre, Donders Institute for Brain, Cognition and Behaviour, The Netherlands
\\
\textbf{5} Computing Sciences, Tampere University, Finland
\\
\textbf{6} Department of Applied Physics, University of Eastern Finland, Kuopio, Finland.
\\
\textbf{7} Institute of Electrical and Biomedical Engineering, UMIT - Private University for Health Sciences, Medical Informatics and Technology, Hall in Tyrol, Austria
\\
\textbf{8} Otto Creutzfeldt Center for Cognitive and Behavioral Neuroscience, University of
M\"unster, Germany
\\
\bigskip

\textcurrency formerly N\"u\ss ing

% Equal Contribution Note
\Yinyang These authors contributed equally to this work.

% Corresponding authorship
* christian.engwer@uni-muenster.de

\end{flushleft}

\section*{Abstract}
Accurate and efficient source analysis in electro- and magnetoencephalography using sophisticated realistic head geometries requires advanced numerical approaches.
This paper presents DUNEuro, a free and open source C++ software toolbox for forward modeling in bioelectromagnetism.
Building upon the DUNE framework, it provides implementations of modern fitted and unfitted finite element methods to efficiently solve the forward problems in electro- and magnetoencephalography.
The user can choose between a variety of different source models that are implemented.
The software's aim is to provide interfaces that are extendible and easy-to-use.
In order to enable a closer integration into existing analysis pipelines, interfaces to Python and Matlab are provided.
The practical use is demonstrated by a source analysis example of somatosensory evoked potentials using a realistic six compartment head model.

\ifarxiv
\else
\linenumbers
\fi

\section*{Introduction}

We present \emph{DUNEuro}, an open source software toolbox for forward modeling in bioelectromagnetism.
Its main focus is to provide an extendible and easy-to-use framework for using various finite element method (FEM) implementations for different neuroscientific applications, such as the electroencephalography (EEG) or magnetoencephalography (MEG) forward problems \cite{Brette2012, Hallez2007, Wolters2007a}.

The general view on a software toolbox can be split into two parts: the user perspective and the developer perspective.
From the perspective of a user, the toolbox should be accessible and easy to use.
Similar methods should work in a similar way through a common interface.
When considering the solution of the EEG forward problem, it should be possible to quickly exchange the discretization scheme.
The user should not be confronted with the high complexity of a \Cpp~finite element code.
Additionally, it should be possible to embed the forward approach into an already existing processing pipeline.
From the point of view of a developer, who wants to implement different discretization schemes or extend already existing approaches, further aspects are important.
As different finite element methods share several subcomponents, such as the representation of the computational domain or the solver of the linear system, the toolbox should bundle the different implementations and enable code reuse.
The toolbox should be extendible, especially with respect to common variable components.
In addition, as there are already several libraries offering codes for finite element computations, it would be advantageous to make use of existing components and benefit from existing maintenance and testing infrastructure.

In order to perform analysis of EEG or MEG data, different open source software packages offer tools to implement a complete processing pipeline such as the Matlab-based toolboxes FieldTrip \cite{Oostenveld2011}, Brainstorm \cite{Tadel2011} and Zeffiro \cite{He2020}, the Python-based toolboxes MNE-Python \cite{Gramfort2013}, FEMfuns \cite{Vermaas2020} or the \Cpp~code MNE \cite{Gramfort2014}.
The toolbox SimBio (\url{https://www.mrt.uni-jena.de/simbio}) offers a broad range for forward and inverse methods for EEG and MEG analysis.
Recently, the EEG FEM forward modeling using the St. Venant source model has been integrated into FieldTrip \cite{Vorwerk2018}.
With respect to the finite element method, only first order Lagrangian elements are implemented in SimBio.
Due to its structure, extending the code to support different discretization schemes would be error-prone and time-consuming.
In addition, there are well-tested and established general purpose software toolboxes for finite element computations.

One existing library for finite element computations is the \emph{distributed and unified numerics environment} DUNE (\url{http://www.dune-project.org}).
More precisely, it is a general purpose open-source \Cpp~library for solving partial differential equations using mesh-based methods \cite{Bastian2020}.
It is extendible by offering a modular structure and providing abstract interfaces and separation between data structures and algorithms.
Due to the modular structure, a user of DUNE only has to use those modules that are needed.
At the core of the DUNE library is an abstract definition of a grid interface \cite{Bastian2008, Bastian2008a}.
Using the abstract interface allows writing reusable code that is independent of the concrete implementation of the grid or the type of the grids elements.
Then the identical code can be used in multiple spatial dimensions and for tetrahedral, hexahedral or other element types.
DUNEuro builds upon several existing DUNE modules such as the dune-uggrid module \cite{Bastian1997} and the dune-pdelab module \cite{Bastian2010}.

Currently, the DUNEuro toolbox is developed and used on Linux operating systems.
The software is made available under the GPL open source license and can be downloaded from the GitLab repository (\url{https://gitlab.dune-project.org/duneuro}), which is also linked from the DUNEuro homepage (\url{http://www.duneuro.org}).

In the following, we first give a short summary on the background of solving the EEG and MEG forward problems with finite element methods.
This is followed by a general description of the toolbox, its use of existing frameworks for solving partial differential equations and its main concepts of different interfaces used within the library.
Afterwards, we will describe a method for localizing elements within a tessellation based on a global coordinate.
The interaction of a user with DUNEuro is done through bindings with a scripting language, which will be presented in the following section.
Subsequently, we use the DUNEuro toolbox to calculate forward solutions which are then used for source analysis on EEG data obtained from a somatosensory experiment.
Finally, a short summary and outlook is given.

\section*{Background}
\label{sec:duneuro-background}

In this section we present the background for solving the EEG and MEG forward problems with finite element methods.

\subsection*{The EEG forward problem}

The aim of the EEG forward problem is to compute the electric potential $u$ in the head domain $\Omega\subset\RR^d$ \cite{Brette2012, Hallez2007, Wolters2007a}.
It can be formulated using Poisson's equation:

\begin{alignat}{2}
  \nabla\cdot\sigma\nabla u &= \nabla\cdot j^p & \quad & \text{in }\Omega \\
  \scp{\sigma\nabla u}{n} &= 0 & & \text{on }\partial\Omega,
  \label{eq:poisson}
\end{alignat}

where $\sigma:\Omega\mapsto\RR^{3\times 3}$ denotes the symmetric and positive definite conductivity tensor, $j^p$ denotes the primary current density and $n$ denotes the unit outer normal on the head surface $\partial\Omega$.
The current source is usually modeled as a current dipole at a position $\dipoleposition\in\Omega$ with the dipole moment $\dipolemoment\in\RR^d$.
With the use of the delta distribution $\delta_{\dipoleposition}$ centered at $\dipoleposition$, the current dipole leads to the source term

\begin{align*}
  \nabla\cdot j^p=M\cdot\nabla\delta_{\dipoleposition}.
\end{align*}

\subsection*{Finite element methods for the EEG forward problem}

Several finite element methods have been proposed to solve the EEG forward problem.
A common basis for these methods is the tesselation of the head domain $\Omega$.
The domain is partitioned into a set of elements of simple shape, such as tetrahedrons or hexahedrons.
Depending on the concrete method, certain regularity assumptions are imposed on the tesselation, such as not containing hanging nodes.
For details on the precise definition of a tesselation we refer to \cite{Brenner2007}.
In the following, we will denote a tesselation of $\Omega$ by $\Tt_h(\Omega)$.
Here, $h\in\RR$ denotes the maximal diameter of a mesh element.

One main difference between several finite element methods is the discrete representation of the potential $u$ and the formulation of the discrete representation of Poisson's equation.
We will not present the mathematical rigorous definition of the different methods but refer to the respective publications that introduced the methods for solving the EEG forward problem.
The conforming finite element method using Lagrangian elements represents the potential as a continuous, piecewise polynomial function \cite{Gencer2004, Lew2009, Medani2015, Pursiainen2011, Schimpf2002, Weinstein2000, Beltrachini2018, Beltrachini2019, Acar2016, Marin1998, VanUitert2004, Vallaghe2010}.
For the discretization of the equation, the classical weak formulation is used directly.
In contrast, the discontinuous Galerkin method does not enforce the continuity of the potential in its function definition but instead, incorporates the continuity weakly through the use of a modified weak formulation \cite{Engwer2017}.
Using this approach, it gains continuity of fluxes on the discrete level.
The mixed finite element method transforms the second order Poisson's equation into a system of first order equation, introducing an additional unknown for the electric field \cite{Vorwerk2017}.

The finite element methods described above have in common that they use a tesselation which resolves the computational domain $\Omega$.
Recently, two finite element methods have been introduced for solving the EEG forward problem that instead use a tesselation of an auxiliary domain $\hat\Omega$ which is independent of $\Omega$.
The conformity of the solution to the domain $\Omega$ is incorporated weakly be modifying the discrete weak formulation.
The CutFEM method uses a function representation that is continuous on each isotropically homogenized tissue compartment \cite{Burman2015, Nuessing2018}.
The unfitted discontinuous Galerkin method additionally transfers the continuity constraint within each subdomain to the weak formulation \cite{Bastian2009, Nuessing2016}.

Due to the difference in representing the discrete function and the different properties of these representations, different strategies for discretizing the dipolar source term have to be taken into account.
In general, it is unclear how to evaluate the derivative of the delta distribution.
Several different approaches for the various finite element methods have been proposed in the literature to handle this singularity.
For the conforming finite element methods, the partial integration approach \cite{Weinstein2000, Yan1991}, the St. Venant approach \cite{Wolters2007, Medani2015, Vorwerk2014}, the full and projected subtraction approach \cite{Wolters2007a, Drechsler2009, Beltrachini2018, Beltrachini2019} and the Whitney approach \cite{Tanzer2005, Bauer2015, Pursiainen2011, Pursiainen2016, Miinalainen2019} have been introduced.
Similar approaches have been presented for the discontinuous Galerkin method, however their exact formulation differs, due to the different discretization approach.
These approaches are the partial integration approach \cite{Vorwerk2016}, the St. Venant approach \cite{Nuessing2018} and the full and localized subtraction approach \cite{Engwer2017, Nuessing2018}.
For the unfitted finite element methods both the partial integration approach and approaches following the principle of St. Venant have been adopted \cite{Nuessing2016, Nuessing2018}.

The discretization of the EEG forward problem leads to a system $Ax=b$ to be solved for $x\in\RR^n$.
In order to speed up the computation of the solution to the EEG forward problem, we first note that in order to compute a lead field matrix, it is not necessary to know the potential in the interior of the domain $\Omega$, but only the evaluation at the sensor positions on the boundary.
For a given source, this leads to a potential vector $U\in\RR^N$, where $N\in\NN$ denotes the number of sensors.
Given a solution $x\in\RR^n$, the evaluation can be represented by a linear map $R\in\RR^{N\times n}$ as $U=Rx$.
Inserting $x=A^{-1}b$ and defining $T:=RA^{-1}$ results in $Tb=U$.
This means, once $T$ is known, the EEG forward problem can be solved by computing the right-hand side vector $b$ and performing a matrix-vector multiplication.
The matrix $T\in\RR^{N\times n}$ is called the (EEG) \emph{transfer matrix}.
By exploiting the symmetry of the discrete operator $A$, the transfer matrix can be computed row-wise by solving $AT^t=R^t$ using the $N$ rows of $R$ as the right-hand sides.
Thus, for the computation of the transfer matrix, the linear system has to be solved once for every sensor location \cite{Weinstein2000, Wolters2004, Gencer2004}.

\subsection*{MEG}

The solution of the MEG forward problem directly follows from the solution
of the respective EEG forward problem via the law of Biot-Savart \cite{Brette2012,Haemaelaeinen1993}.
For a current distribution $j$, the magnetic field B at a position $y\in\RR^d$ is computed via

\begin{align}
  B(y) = \frac{\mu_0}{4\pi}\int_{\Omega} j(x) \times \frac{y-x}{\| y-x\|^{3}} \dd{x},
  \label{eq:meg}
\end{align}

where $\times$ denotes the three-dimensional cross product and $\mu_0$ the permeability of free space.

By splitting the current distribution $j$ into the primary current $j^p$ and secondary current $j^s=-\sigma\nabla u $, the magnetic field B in equation (\ref{eq:meg}) can be divided into the primary magnetic field $B^p$ and the secondary magnetic field $B^s$.
While there is an analytical expression for $B^p$ \cite{Brette2012,Haemaelaeinen1993}, in order to compute $B^s$ the integral expression

\begin{align*}
  \int_{\Omega} \sigma(x)\nabla u(x) \times \frac{y-x}{\| y-x\|^{3}} \dd{x}
\end{align*}

needs to be computed numerically.
For the standard continuous Galerkin method (CG-FEM) this integral can be directly evaluated using the discrete representation of the potential $u_h$.
Results for the MEG approach for the discontinuous Galerkin method (DG-FEM) in \cite{Piastra2018} indicate that a direct usage of $\sigma\nabla u_h$ leads to suboptimal accuracies.
Instead, the numerical flux of the discontinuous Galerkin method should be used, see \cite{Piastra2018,Piastra2019}.
Similarly to the EEG case, a transfer matrix can be derived which allows computing $B^s$ at the sensors with a matrix-vector multiplication, instead of solving the EEG forward problem and computing the integral subsequently, see\cite{Wolters2004,Piastra2019}.
Note that when using the subtraction approach, the resulting solution does not include the contributions of the singularity potential.

\section*{Library interfaces}
\label{sec:duneuro-library}

DUNEuro builds upon several existing DUNE modules:
For representing geometry-conforming tetrahedral and hexahedral meshes, we use the grid implementations provided by the dune-uggrid module \cite{Bastian1997}.
In order to reduce the memory consumption and to simplify the user-code when using a geometry-adapted hexahedral mesh \cite{Wolters2007}, we use the dune-subgrid module to extract parts of a mesh that is given as a segmented voxel image \cite{Graeser2009}.
The discretization of the partial differential equation makes use of the dune-pdelab module \cite{Bastian2010}.
In dune-pdelab, many different discretization schemes along with appropriate finite elements are implemented allowing a rapid prototyping of new models.
It offers abstractions for the concept of a function space on a grid or for the linear operator used in the discretization.
The implementation of the unfitted discontinuous Galerkin method is provided in the dune-udg module \cite{Engwer2012}.
One component of the unfitted finite element methods is the integration over implicitly defined domains, which is performed using the \Cpp~library tpmc (\url{http://github.com/tpmc}).
For the solution of the linear system, we make use of the \emph{iterative solver template library (ISTL)} offered by the dune-istl module \cite{Blatt2006}.

In this section we present in detail several subcomponents and interfaces of the DUNEuro toolbox and give information on the extendibility of each component. We describe the driver interface, the discretization of the forward model and the implemented source models.

\subsection*{The EEG-MEG driver interface}

As described above, there are several different discretization schemes available for solving the EEG forward problem and each scheme provides different source models.
The finite element methods presented here can be split into two different categories: the \emph{fitted} and \emph{unfitted} discretization methods.
The fitted category refers to a discretization method that uses a grid whose geometry is fitted to the model geometry.
The basis of this approach is a \lstinline{VolumeConductor} class that stores the grid along with the conductivity tensor of each grid element.
Currently, there are two different fitted discretization schemes implemented in DUNEuro: the conforming Galerkin (CG-FEM) and the discontinuous Galerkin (DG-FEM) finite element methods.
Methods that fall into this category but are not yet available are mixed finite element methods or finite volume schemes.
The unfitted category refers to a discretization method that uses a grid which is independent of the model geometry and employs the model geometry weakly.
The model geometry is provided implicitly via level-set functions and considered in the weak formulation.
Currently, the unfitted discontinuous Galerkin (UDG) method is implemented as a discretization scheme in the unfitted category.

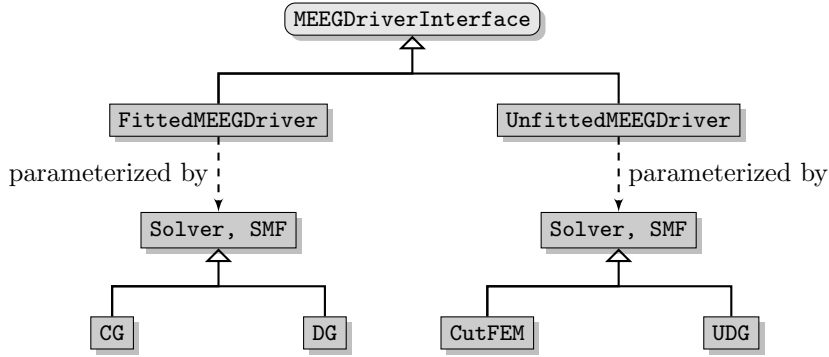
\begin{figure}[!h]
  \centering
  \tikzstyle{arrowimplements}=[->, >=open triangle 90, thick]
\tikzstyle{arrowparameterized}=[->, >=latex', dashed, thick]
\tikzstyle{line}=[-, thick]
\tikzstyle{interface}=[rectangle, draw=black, rounded corners, fill=gray!20, drop shadow, text centered, anchor=north, text=black]
\tikzstyle{implementation}=[rectangle, draw=black, fill=gray!40, drop shadow, text centered, anchor=north, text=black]
\begin{tikzpicture}[node distance=1]
  \node (driverinterface) [interface] {\lstinline{MEEGDriverInterface}};
  \node (auxnode0) [below=of driverinterface] {};
  \node (fitteddriver) [implementation, left=of auxnode0] {\lstinline{FittedMEEGDriver}};
  \node (unfitteddriver) [implementation, right=of auxnode0] {\lstinline{UnfittedMEEGDriver}};
  \node (ssmffitted) [implementation, below=of fitteddriver] {\lstinline{Solver, SMF}};
  \node (ssmfunfitted) [implementation, below=of unfitteddriver] {\lstinline{Solver, SMF}};
  \node (auxnode1) [below=of ssmffitted, anchor=center] {};
  \node (cg) [implementation, left=of auxnode1] {\lstinline{CG}};
  \node (dg) [implementation, right=of auxnode1] {\lstinline{DG}};
  \node (udg) [implementation, below=of ssmfunfitted, anchor=center] {\lstinline{UDG}};
  \draw[arrowimplements] (fitteddriver.north) -- ++(0,0.4) -| (driverinterface.south);
  \draw[line] (fitteddriver.north) -- ++(0,0.4) -| (unfitteddriver.north);
  \draw[arrowparameterized] (fitteddriver.south) -- node [left,midway] {parameterized by} (ssmffitted.north);
  \draw[arrowparameterized] (unfitteddriver.south) -- node [right,midway] {parameterized by} (ssmfunfitted.north);
  \draw[arrowimplements] (cg.north) -- ++(0,0.4) -| (ssmffitted.south);
  \draw[line] (cg.north) -- ++(0,0.4) -| (dg.north);
  \draw[arrowimplements] (udg.north) -- ++(0,0.4) -| (ssmfunfitted.south);
\end{tikzpicture}
  \caption{{\bf DUNEuro driver interface diagram.}
  Diagram showing the structure of the driver interface and its implementations. The \lstinline{SourceModelFactory} is abbreviated as \lstinline{SMF}.}
  \label{fig1}
\end{figure}

From the user perspective of a software framework, it should be simple and intuitive to change from a fitted to an unfitted discretization or between different discretization schemes within each category.
For example switching from CG-FEM to DG-FEM should not require fundamental changes in the user code.
A further consideration when designing the interface of the software is the way the user will interact with it.
As described in more detail below, we want to provide bindings to languages a potential user is already familiar with, such as Python or Matlab.
In order to simplify both, the overall user interface as well as the process of creating such bindings, we define a single coarse grained interface class to interact with the internal toolbox.
This interface class is called the \lstinline{MEEGDriverInterface}.
It describes the general concepts of solving EEG and MEG forward problems.
Each of the two discretization categories is implemented by its own driver class, the \lstinline{FittedMEEGDriver} and the \lstinline{UnfittedMEEGDriver} respectively.
Fig~\ref{fig1} shows a general diagram of the \lstinline{MEEGDriverInterface}.
For each category, the implementation of the discretization scheme is provided via two template parameters: a \lstinline{Solver} and a \lstinline{SourceModelFactory}.
The purpose of the solver class is to bundle the handling of the system matrix and the solution of the resulting linear system.
The source model factory will construct source models whose purpose is the assembly of the right-hand side.
Both, the solver and the source model factory, are further described below.
The user of the toolbox will not directly interact with the implementation of the drivers, but only with the driver interface class.

\subsection*{The solver and the source model factory}

\label{sec:duneuro-discretization}
The purpose of the solver class is the assembly of the system matrix and the solution of the linear system.
It contains the discretization scheme as well as the necessary function spaces for representing discrete functions.
The main interface method is a \lstinline{solve} method which, given a right-hand side vector, solves Poisson's equation and returns the discrete solution.
Several forward problems in bioelectromagnetism, e.g., the EEG forward problem, electric \cite{Opitz2015, Wagner2016} or magnetic stimulation \cite{Opitz2011, Janssen2013} or the computation of a transfer matrix, mainly differ with respect to the right-hand side of the linear system.
The solver class can thus be reused for any such purpose.
By using a single solver class, the system matrix has to be assembled only once and can be reused for further purposes.
As the different discretization schemes differ in the way the matrix is stored, e.g., with respect to the blocking scheme of the matrix entries, this information is hidden from the interface.
The purpose of the source model factory is to construct the different source models dynamically based on a configuration provided by the user.
All source models provide a common interface which is described below.

We will illustrate the extendibility with respect to the discretization scheme using the example of a mixed finite element method.
MixedFEM is based on a first order representation of Poisson's equation and employs unknowns for both, the potential and the electric field \cite{Vorwerk2017}.
It derives a weak formulation and uses scalar and vector-valued finite elements on a geometry conforming grid as a discretization.
It thus falls into the category of fitted discretization schemes.
In order to use the described \lstinline{FittedMEEGDriver}, one needs to provide two components: a \lstinline{MixedFEMSolver} and a \lstinline{MixedFEMSourceModelFactory}.
The \lstinline{MixedFEMSolver} contains the discretization of the stiffness matrix as well as the definition to solve the resulting linear system.
The implementation of such a solver class is heavily based on the \lstinline{dune-pdelab} module which contains, for example, the implementations of the local basis functions.
The \lstinline{MixedFEMSourceModelFactory} offers a method to create different source models for the MixedFEM approach, whose purpose is then to assemble the right-hand side vector for a given source position.
In \cite{Vorwerk2017}, two different source models have been presented: a \emph{direct approach} and a \emph{projected approach}.
Finally, one has to provide means to evaluate a discrete solution at electrode positions along with the resulting right-hand side of the transfer matrix approach.
Once these components are implemented, the features of the driver, e.g., computing a transfer matrix or solving the EEG forward problem, are available.

\subsection*{Source models}

For each discretization method, there are several different source models that are used to discretize the mathematical point dipole.
A list of source models currently supported by DUNEuro for different FEM discretizations is provided in Table \ref{table1}.

\begin{table}[!ht]
  %\begin{adjustwidth}{-2.25in}{0in} % Comment out/remove adjustwidth environment if table fits in text column.
  \centering
  \caption{\bf Overview of source models currently supported for EEG/MEG by DUNEuro for different FEM discretization schemes.}
  \begin{tabular}{lll|l|l|l|}
    && \multicolumn{4}{c}{\bf Source models}\\
    \cline{3-6}
    &&\multicolumn{1}{ |c| }{\bf Partial}& \multicolumn{1}{ |c| }{\bf St. Venant}& \multicolumn{1}{ |c| }{\bf Subtraction} & \multicolumn{1}{ |c| }{{\bf Whitney}\textsuperscript{1}}\\
    &&\multicolumn{1}{ |c| }{{\bf integration} \cite{Bauer2015}} &  \multicolumn{1}{ |c| }{\cite{Bauer2015}}& \multicolumn{1}{ |c| }{\cite{Wolters2007a,Drechsler2009}}  & \multicolumn{1}{ |c| }{\cite{Miinalainen2019}} \\
    \cline{2-6}
    \multirow{6}{*}{\rotatebox[origin=c]{90}{\bf FEM}}
    & \multicolumn{1}{ |c| }{\bf CG} & EEG/MEG & EEG/MEG & EEG/MEG & EEG/MEG\\
    & \multicolumn{1}{ |c| }{\cite{Vorwerk2014}}&&&& \\
    \cline{2-6}
    & \multicolumn{1}{ |c| }{\bf DG} & EEG/MEG\textsuperscript{2}& - & EEG/MEG\textsuperscript{2}&- \\
    & \multicolumn{1}{ |c| }{\cite{Engwer2017,Piastra2018}}&&&& \\
    \cline{2-6}
    & \multicolumn{1}{ |c| }{\bf UDG}& EEG & - & EEG &- \\
    & \multicolumn{1}{ |c| }{\cite{Nuessing2016}}&&&& \\
    \cline{2-6}
  \end{tabular}
  \begin{flushleft} \textbf{1} The Whitney source model is currently only implemented for tetrahedral meshes.\\
  \textbf{2} In the MEG implementation, the numerical flux for the secondary magnetic field for DG-FEM is currently only implemented for hexahedral meshes.
  \end{flushleft}
  \label{table1}
  %\end{adjustwidth}
\end{table}

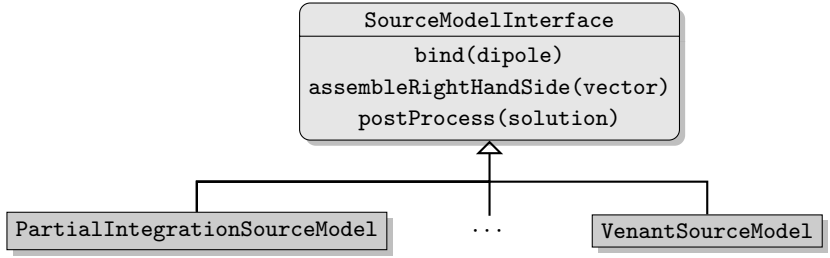
\begin{figure}[!h]
  \centering
  \tikzstyle{arrowimplements}=[->, >=open triangle 90, thick]
\tikzstyle{arrowparameterized}=[->, >=latex', dashed, thick]
\tikzstyle{line}=[-, thick]
\tikzstyle{interface}=[rectangle, draw=black, rounded corners, fill=gray!20, drop shadow, text centered, anchor=north, text=black]
\tikzstyle{implementation}=[rectangle, draw=black, fill=gray!40, drop shadow, text centered, anchor=north, text=black]
\begin{tikzpicture}[node distance=1]
  \node (sourcemodelinterface) [interface,rectangle split, rectangle split parts=2, align=center]
  {
  \lstinline{SourceModelInterface}
  \nodepart{second}
  \lstinline{bind(dipole)}\\
  \lstinline{assembleRightHandSide(vector)}\\
  \lstinline{postProcess(solution)}
  };
  \node (dottedsm) [below=of sourcemodelinterface] {$\cdots$};
  \node (pism) [implementation, left=of dottedsm] {\lstinline{PartialIntegrationSourceModel}};
  \node (venantsm) [implementation, right=of dottedsm] {\lstinline{VenantSourceModel}};
  \draw[arrowimplements] (pism.north) -- ++(0,0.4) -| (sourcemodelinterface.south);
  \draw[line] (pism.north) -- ++(0,0.4) -| (venantsm.north);
  \draw[line] (pism.north) -- ++(0,0.4) -| (dottedsm.north);
\end{tikzpicture}
  \caption{{\bf DUNEuro source model interface diagram.}
  The structure of the source model interface and its implementations.}
  \label{fig2}
\end{figure}

The common task of source models can be stated as: given a dipole position and a dipole moment, assemble the right-hand side vector.
This right-hand side vector will then be passed on to the respective solver class described above.
As there is still research ongoing and new source models are being developed, it should be easy to provide an additional source model without having to modify the existing code.
In addition, it should be possible to choose the source model at runtime, both for investigating the effects of different source models as well as ruling out the source model as a source of errors.
Some source models, such as the subtraction approach, do not provide a right-hand side for the full potential, but need to apply an additional post-processing step to the resulting solution in order to obtain the full potential.
For the subtraction approaches, this post-processing step consists of adding the singularity potential to the correction potential.
As this post processing step depends on the type of the source model and the user should have the option to turn off the post-processing, it is provided as a method of the source model interface.
Fig~\ref{fig2} shows a diagram of the general \lstinline{SourceModelInterface} along with its implementations.

A main advantage of the direct source models such as the partial integration approach or the St. Venant approaches, is the sparsity of the right-hand side.
When stored inside a sparse vector container, the time for applying the transfer matrix, i.e., multiplying the right-hand side with the transfer matrix, can be reduced.
The complexity is $\Oo(M)$, where $M$ denotes the number of mesh elements.
The constant is proportional to the number of non-zero entries of the right-hand side.
The latter is usually independent of the mesh resolution.
For indirect source models such as the subtraction approach, using the same data type as for the sparse source models would introduce an additional overhead.
Thus, in order to be able to handle dense and sparse vector types for the right-hand side, the vector type is provided as a template parameter of the source model interface.

We will illustrate the extendibility with respect to the source models on the example of a modified subtraction approach for CG-FEM.
In \cite{Nuessing2018} a modification of the subtraction approach has been presented: the \emph{localized subtraction approach}.
It restricts the contribution of the singularity part of the potential to a patch around the source location.
As the functions within a DG-FEM discretization can be discontinuous, they can directly capture the jump occurring at the boundary of the patch.
For a CG-FEM discretization, such jumps can not be directly resolved and thus the localization scheme has to be modified.
Instead of using a restriction of the singularity contribution to the patch, one can multiply the singularity contribution with a function that linearly interpolates within an interface zone of the patch between the singularity potential and zero.
A source model implementing this localized subtraction approach would provide a class fulfilling the source model interface.
Within the \lstinline{bind} method, the local patch would be created and the linear interpolation in the interface zone could be constructed.
The implementation of the \lstinline{assembleRightHandSide} method contains the integration of the different model terms, resulting in the right-hand side.
The \lstinline{postProcess} method adds the singularity potential to the correction potential on the local patch.

\section*{Element localization}
\label{sec:duneuro-localization}

A common subtask when assembling the right-hand side for a given dipole is the localization of the mesh element containing the dipole.
For a sparse source model, this is especially relevant, as the time of assembling the right-hand side is usually constant, once the dipole element has been found.
The complexity of the right-hand side assembly thus strongly depends on the complexity of the method that is used for finding the dipole element.
The most straightforward approach is given by a linear search among the mesh elements.
Assuming an ordering of the mesh elements, we evaluate for each element of the mesh if it contains the dipole position.
Once the result of the evaluation is positive, we return this element.
This algorithm has an average and worst-case complexity of $\Oo(M)$, where $M$ denotes the number of the mesh elements.

A first step to speed up the localization can be found by using geometric information when iterating the mesh elements instead of using a fixed ordering \cite{Brown1997}.
The method presented here is called \emph{edge hopping}:
\begin{enumerate}
\item Start at a given mesh element and iterate over all faces of the current element. \label{it:eh_start}
\item Compute the relative position of the dipole location and the hyperplane induced by the face center and its outer normal.\label{it:eh_face}
\item If the dipole lies in normal direction, continue the search at step \ref{it:eh_start} with the neighboring element if such an element exists.
\item If the face has no neighboring element, the dipole lies outside of the mesh or the mesh is not convex. Terminate the search.
\item If the dipole lies in the opposite direction, continue at step \ref{it:eh_face} with the evaluation of the next face.
\item If the dipole lies on the inside of all faces of the current element, the dipole element has been found.
\end{enumerate}
A requirement of the edge-hopping method is the convexity of the mesh, that is usually only fulfilled by the multi-layer sphere models, and not by the realistically shaped head models.
However, as the algorithm monotonously moves closer to the dipole element, we only need convexity of the mesh in a sphere around the dipole location and the starting point of the iteration.
As the considered sources lie in the gray matter compartment, that is completely enclosed by the skin, we can easily find such a sphere around the source locations if the starting location is close to the source position.
In order to find an element that is close to the source location, we insert the element centers into a \emph{k-d Tree}, that is a data structure to efficiently perform nearest neighbor searches \cite{Bentley1975}.
It does so by recursively splitting the set of element centers along the Cartesian directions.
Even though the center of the element which is closest to the dipole location does not have to belong to the element containing the dipole, it can be assumed to be close to the desired element.
It thus offers an efficient starting point for the edge-hopping algorithm.

\section*{Interface to scripting languages}
\label{sec:duneuro-scripting}

In this section we describe the interaction of a user with the DUNEuro library.
In general, a common approach is to provide a compiled binary executable that the user is able to call directly.
This executable would then load the data provided by the user from the hard disk, perform the desired computation and write the computed result back to the hard disk.
As different users might want to perform different sets of computations, the computations to be performed can be configured by the user, either through command line parameters or through a configuration file.
An advantage of this approach is its very simple and straightforward usage, similar to any other executable on the operating system.
There is no need for additional packages or additional software and the executable can be used directly by the user.
However, the computation of the solution to the forward problems is usually only a small part in a longer pipeline for source analysis.
This pipeline usually consists of the data measurements and pre-processing steps and the forward solution is part of an inverse estimation process.
When using the library directly in an executable, one has to provide methods for reading any input data as well as writing out the resulting output.
Similarly, the configuration has to be transferred to the executable by the user.

A more convenient way to use the provided library can be found by offering bindings to a scripting language such as Matlab (The Math Works Inc., Natick, Massachusetts, United States; \url{https://www.mathworks.com}) or Python (Python Software Foundation, \url{https://www.python.org}).
For both languages there are already existing software frameworks for processing EEG and MEG data \cite{Gramfort2013,Oostenveld2011,Tadel2011, He2020}.
Thus, by providing direct bindings one can include the forward modeling approach directly into an existing analysis pipeline.
An example for such an integration is presented in \cite{Vorwerk2018}, where the authors introduce a pipeline for performing EEG source analysis using the conforming finite element method together with the classical St. Venant source model.
The forward models are implemented using the SimBio software (\url{https://www.mrt.uni-jena.de/simbio}) and integrated into the Matlab-based FieldTrip-toolbox \cite{Oostenveld2011}.

\begin{figure}[!h]
  \centering
  \scalebox{.9}{
\begin{tikzpicture}[node distance=3pt,
  defaultblock/.style={
    draw=white,
    font={\bfseries},
    align=center,
    text height=12pt,
    text depth=6pt,
    anchor=north west,
  },
  blueb/.style={defaultblock,
    text=white,
    fill=gray,
    rounded corners,
    text width=2cm,
    },
  grayb/.style={blueb,fill=black!60!orange},
  greenb/.style={blueb,fill=black!60!green},
  orangeb/.style={blueb,fill=black!20!orange},
  blueb2/.style={blueb,fill=gray},
  grayb2/.style={blueb2,fill=gray},
  orangeb2/.style={blueb2,fill=black!20!white},
  ]
  \node[defaultblock](A){};
  \node[defaultblock,below=of A](B){};
  \node[defaultblock,below=of B](C){};
  \node[blueb,right=of A](FT) {FieldTrip};
  \node[blueb2,right=of FT](BS) {Brainstorm};
  \node[defaultblock,right=of BS,text width=1cm](X1) {\Large$\cdots$};
  \node[blueb2,right=of X1] (nipy) {nipy/MNE};
  \node[defaultblock,right=of nipy,text width=1cm](X2) {\Large$\cdots$};
  \node[grayb,right=of B,text width=5cm+20pt] (dnmat) {DUNEuro-matlab};
  \node[grayb,right=of dnmat,text width=5cm+20pt] (dnpy) {DUNEuro-py};
  \node[greenb,below=of dnmat.south west,anchor=north west,text depth=34pt,text width=2.5cm]  (matlab) {MATLAB};
  \node[grayb,right=of matlab.north east,anchor=north west,text width=5cm+30pt] (dn) {DUNEuro};
  \node[orangeb,below=of dn,text width=5cm+30pt] (dune) {DUNE};
  \node[greenb,right=of dn.north east,anchor=north west,text depth=34pt,text width=2.5cm] (PY) {Python};
\end{tikzpicture}}
  \caption{{\bf Modular structure.}
  Relation of the DUNEuro modules with respect to DUNE, external software and downstream libraries.}
  \label{fig3}
\end{figure}
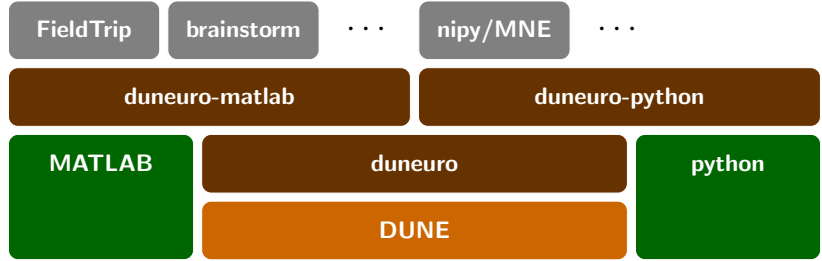

As it builds upon several existing modules of the software toolbox DUNE, the core functionality is provided by the \Cpp module DUNEuro.
Bindings to the Matlab and Python scripting languages are provided in separate DUNE modules: \emph{DUNEuro-py} and \emph{DUNEuro-matlab}, respectively.
A structural overview of DUNEuro and its interface modules with respect to DUNE, external software and downstream libraries is illustrated in Fig~\ref{fig3}.
The purpose of both modules is to translate the input data given as data structures in the respective programming language and translate them into the \Cpp~counterparts.
For some cases, this translation can be performed without copying any data, which is especially relevant for large matrices such as the transfer matrix.
An example of the driver construction in a Python script is shown in Listing \ref{lst:basic_python}.
\begin{lstfloat}[!h]
  \caption{Example Python script for creating an \lstinline{MEEGDriver}.}\label{lst:basic_python}
\begin{lstlisting}[style=pythoncode]
import duneuropy as dp
config = {
  'type' : 'fitted',
  'solver_type' : 'cg',
  'element_type' : 'tetrahedron',
  'volume_conductor' : {
    'grid.filename' : 'path/to/grid.msh',
    'tensors.filename' : 'path/to/tensors.dat'
  }
}
driver = dp.MEEGDriver3d(config)
\end{lstlisting}
\end{lstfloat}
The configuration of the discretization is provided as a Python dictionary and the mesh is loaded from a file.
Alternatively, the mesh can also be provided directly by specifying the vertices, elements, labels and conductivity tensors.
Note that the discretization method, in this case \lstinline{cg}, is provided as a parameter in the configuration.
By changing it to \lstinline{dg} and adding the necessary additional parameters such as the penalty parameter $\eta$, one can directly use the discontinuous Galerkin method through the same interface.
Thus, once a user is able to use the DUNEuro library for any discretization method, a switch to a different discretization method can be directly performed.
Listing \ref{lst:basic_matlab} shows the same construction of the driver object as in Listing \ref{lst:basic_python} using the Matlab interface.
\begin{lstfloat}[!h]
  \caption{Example Matlab script for creating an \lstinline{MEEGDriver}.}\label{lst:basic_matlab}
\begin{lstlisting}[style=matlabcode]
cfg = [];
cfg.type = 'fitted';
cfg.solver_type = 'cg';
cfg.element_type = 'tetrahedron';
cfg.volume_conductor.grid.filename = 'path/to/grid.msh';
cfg.volume_conductor.tensors.filename = 'path/to/tensors.dat';
driver = duneuro_meeg(cfg);
\end{lstlisting}
\end{lstfloat}
The general structure of the Matlab script is similar to the Python script.
The main differences are the use of Matlab syntax and the replacement of the Python dictionary by a Matlab struct array.
Even though the wrapper code for creating the driver object is different, both scripting languages interface the \Cpp~library and use the same code base.

\section*{Example: source analysis of somatosensory evoked potentials}
\label{sec:duneuro-practical}

\begin{figure}[!h]
  \centering
  \begin{subfigure}{.6\textwidth}
    \centering
    \includegraphics[width=\textwidth]{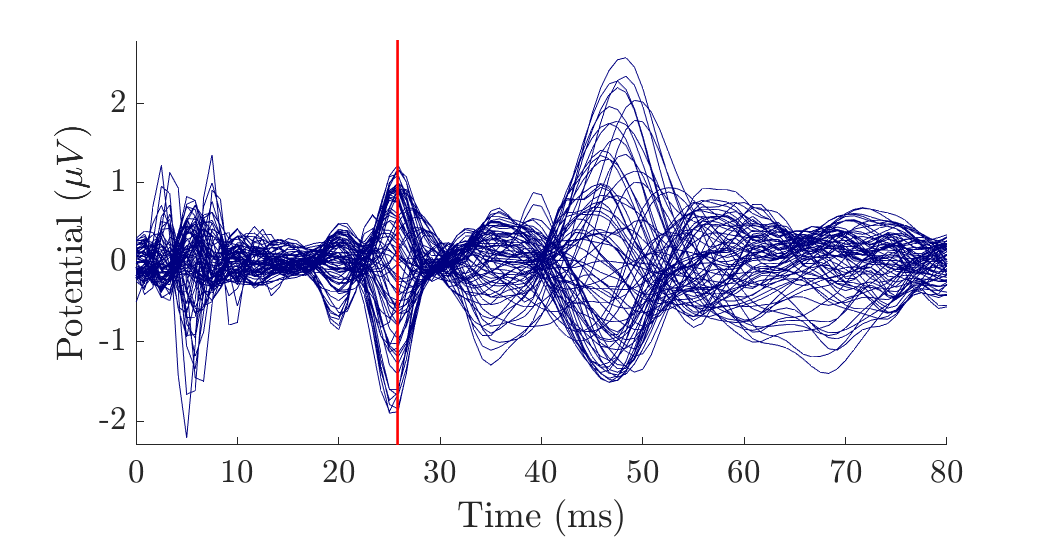}
  \end{subfigure}\hfill{}
  \begin{subfigure}{.38\textwidth}
    \centering
    \includegraphics[height=3.5cm]{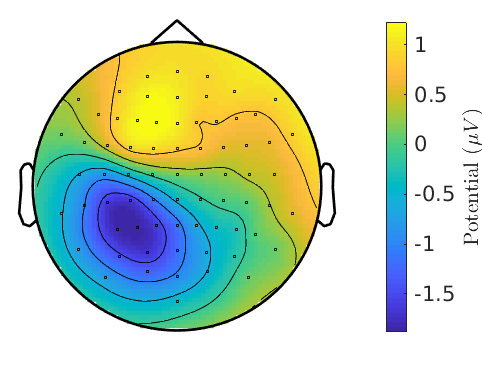}
  \end{subfigure}\hfill{}
  \caption{{\bf Practical application: Preprocessing.}
  Left: Butterfly plot of the somatosensory evoked potentials. The vertical red line indicates the \SI{25.8}{\ms} time point. Right: topography plot of the averaged potential at the electrodes for $t=\SI{25.8}{\ms}$}
  \label{fig4}
\end{figure}

As a practical example the forward solutions of DUNEuro are used to perform a dipole scan on somatosensory evoked potentials using the UDG finite element method.
A right-handed, 49 years old, male subject participated in a somatosensory evoked potential (SEP) EEG recording of an electric stimulation of the right median nerve, available from \cite{Piastra2020}.
The subject gave its written informed consent and all measurements have been approved by the ethics committee of the University of Erlangen, Faculty of Medicine on 20.02.2018 (Ref No 4453 B).
The EEG was measured using 74 electrodes, whose positions where digitized using a Polhemus device (FASTRAK, Polhemus Incorporated, Colchester, Vermont, USA).
The subject was stimulated in supine position in order to reduce modeling errors due to brain movement, because the corresponding MRI for head volume conductor modeling was also measured in supine position \cite{Rice2013}.
In total, 1200 stimuli were applied, each with a duration of \SI{200}{\ms}.
The inter-stimulus interval was randomized in the range of \SIrange{350}{450}{\ms}.
The EEG data was preprocessed using FieldTrip \cite{Oostenveld2011} with a band-pass filter from \SIrange{20}{250}{\hertz} and notch-filters at \SI{50}{\hertz} and harmonics to reduce power-line noise.
After removing one bad channel (P7) the remaining trials where averaged to produce the evoked potential data.
Fig~\ref{fig4} shows a butterfly plot of the resulting time series of the averaged potentials as well as a topography plot of the potential measured at the electrodes at the peak of the P20 component at the time point of \SI{25.8}{\ms} due to the time delay until the stimulus arrives at the median nerve.

\begin{figure}[!h]
  \centering
  \includegraphics[height=6cm]{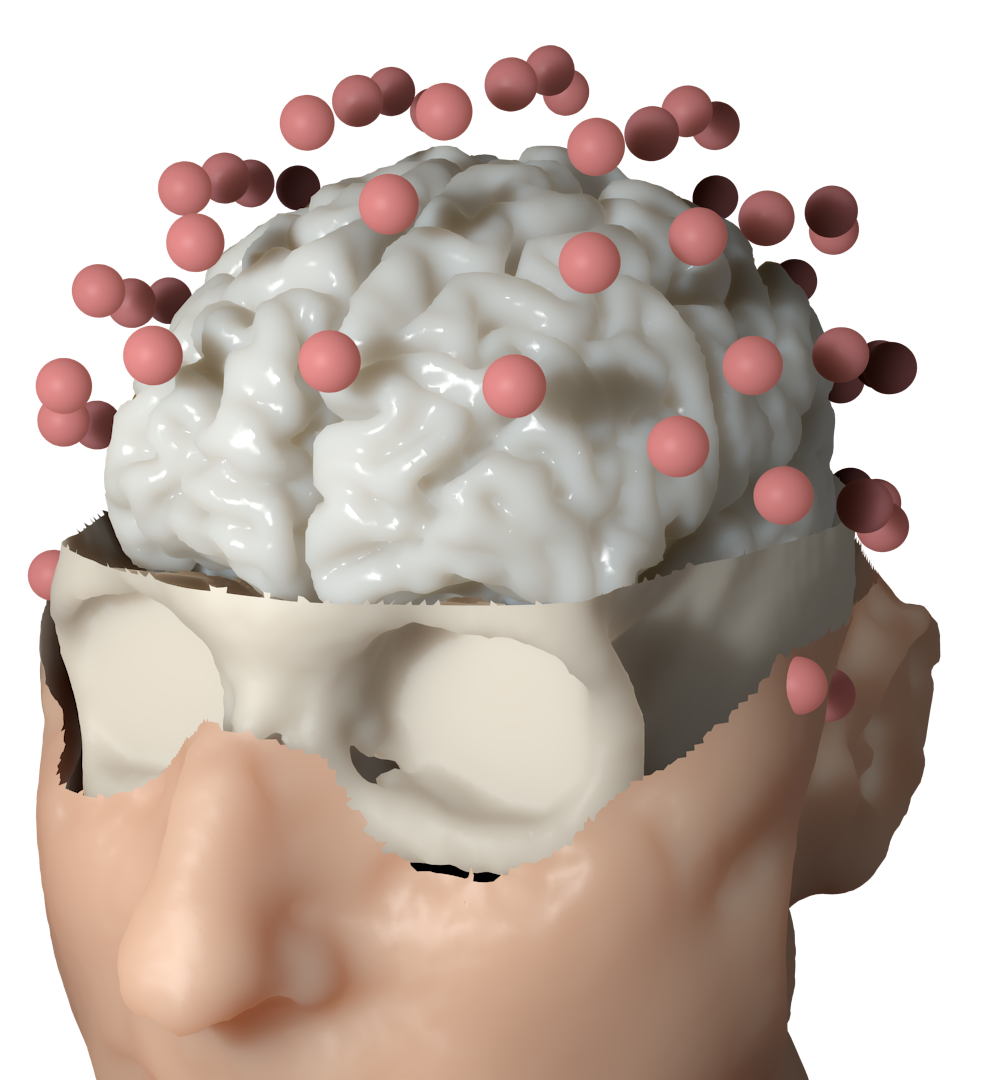}
  \includegraphics[height=6cm]{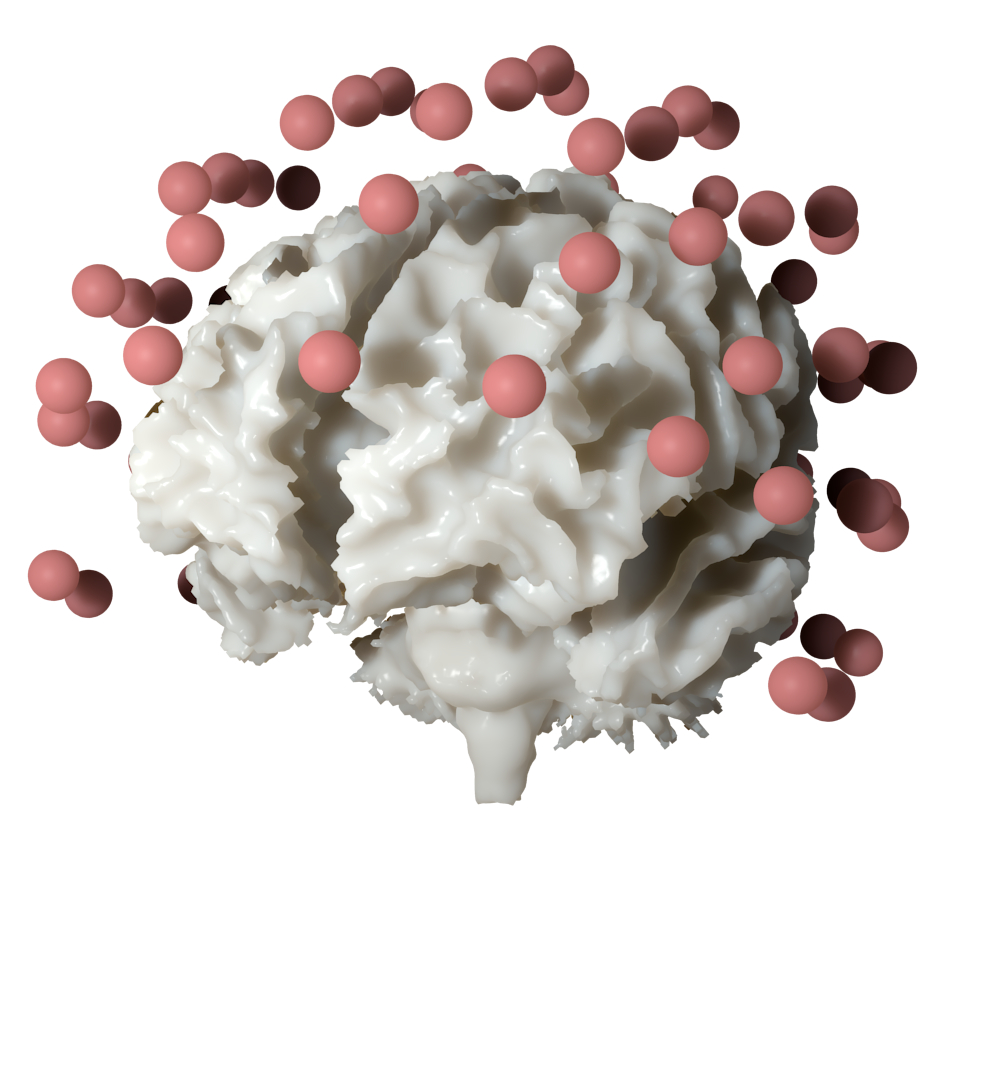}
  \caption{{\bf Practical application: Realistic head model.}
  Left: skin, skull and gray matter surfaces of the six-compartment isotropic head model along with the electrode montage used in the practical example. Right: source space relative to the electrode positions.}
  \label{fig5}
\end{figure}

Using a \SI{3}{\tesla} MRI scanner (Siemens Medical Solutions, Erlangen, Germany), T1-weighted and T2-weighted MRI sequences were measured.
Based on these MRI images, a six-compartment voxel segmentation has been constructed, distinguishing between skin, skull compacta, skull spongiosa, csf, gray matter and white matter using SPM12 (\url{http://www.fil.ion.ucl.ac.uk/spm/software/spm12}) via FieldTrip \cite{Oostenveld2011}, FSL \cite{Jenkinson2012} and internal Matlab routines.
We extracted surfaces from this voxel segmentation to distinguish between the different tissue compartments.
To smooth the surfaces while maintaining the available information from the voxel segmentation, we applied an anti aliasing algorithm created for binary voxel images presented in \cite{Whitaker2000}.
The resulting smoothed surfaces are represented as level-set functions and are available from \cite{Piastra2020}.
The digitized electrodes were registered to the head surface using landmark-based rigid parametric registration.
Especially in occipital and inferior regions, due to the lying position of the subject during MRI measurement, the gray matter compartment touches the inner skull surface.
From Fig~\ref{fig4} we see a clear dipolar pattern in the topography plot.
To estimate the location of the dipole, we performed a single dipole deviation scan using a normal-constraint for dipole orientation on the source space, i.e., a set of source locations within the gray matter compartment \cite{Haemaelaeinen1993}.
The source space is created using a weighted sum of level-set functions for gray and white matter as $\alpha\Phi_{\text{wm}}+(1-\alpha)\Phi_{\text{gm}}$ with $\alpha=0.8$.
The resulting level-set function for the source space was discretized using the marching cubes algorithm presented in \cite{Engwer2017a}, which resulted in \num{256134} source locations.
For each location, we computed the dipole orientation normal to the surface of the source space.
Fig~\ref{fig5} shows the skin, skull and gray matter surfaces and the electrode positions as well as the source space that was used in the example computation.

Using the level-set functions, we constructed an unfitted FEM head model and computed the EEG transfer matrix for all electrode positions using the presented DUNEuro toolbox.
With this transfer matrix, we computed the EEG forward solution for all dipole positions with the fixed orientation and unit strength using the partial integration source model \cite{Weinstein2000, Vorwerk2016}.
The optimal strength $s$ with respect to a given measurement $m$ for a dipole with the leadfield $l$ can be obtained by minimizing $\|ls-m\|_2$ over $s$.
The resulting optimal strength for reproducing the measured data is given as

\begin{align*}
  s = \max\left(\frac{\scp{l}{m}_2}{\|l\|^2_2},0\right).
\end{align*}

The maximum with 0 is used in order to restrict the solution along the respective positive normal direction because the P20/N20 component is assumed to be located in Brodmann area 3b of the somatosensory SI cortex pointing out of the cortex to produce a frontal positivity \cite{Aydin2014, Antonakakis2020}.
This strength is embedded into the goodness of fit measure (GOF) that is defined as

\begin{align*}
  \operatorname{GOF}=1-\frac{\|ls-m\|_2^2}{\|m\|_2^2}
\end{align*}

and measures the ability of the numerical solution to reproduce the measured data.
If the data can be exactly reproduced, the GOF has a value of 1.
In the case of a single dipole deviation scan this includes how well the data can be represented as a single dipole.
Former results have shown that the single dipole source model is appropriate for the reconstruction of the early somatosensory response \cite{Allison1991, Nakamura1998, Aydin2014, Antonakakis2020}.

\begin{figure}[!h]
  \centering
  \includegraphics[height=4.5cm]{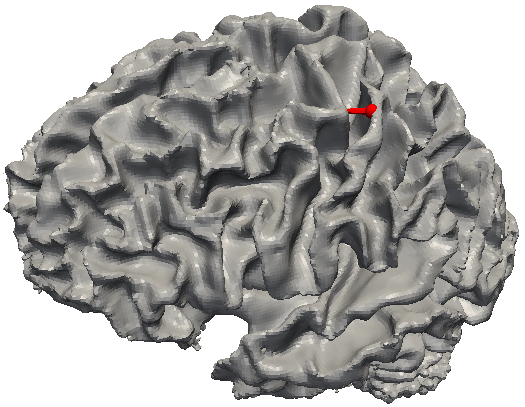}
  \includegraphics[height=4.5cm]{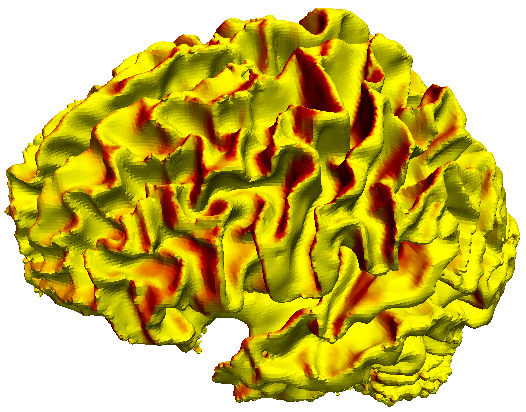}
  \includegraphics[height=4.5cm]{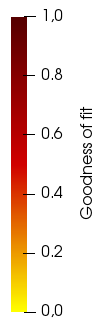}
  \caption{{\bf Practical application: Source reconstruction.}
  Left: Reconstructed source (red) of the P20 component and its location within the source space. Right: Distribution of the GOF measure on the source space. A darker color indicates a higher GOF.}
  \label{fig6}
\end{figure}

The source was reconstructed in the primary somatosensory cortex in the wall of the post-central gyrus with a GOF of 0.962 and with a mainly tangential orientation, which reproduces findings of \cite{Allison1991, Nakamura1998, Aydin2014, Antonakakis2020}.
Fig~\ref{fig6} shows the source embedded in the source space and the distribution of the GOF measure.
We see that the GOF measure is higher for source locations on the gyral walls with a tangentially oriented normal vector and that the higher values are located close to the central sulcus.
Overall, the GOF measure shows a smooth distribution in these areas while being sensitive to orientation changes.

\section*{Summary and outlook}
\label{sec:duneuro-conclusion}

In this paper we presented the DUNEuro software, a toolbox for solving forward problems in bioelectromagnetism.
We provided a general description of the toolbox as well as detailed information about the main concepts.
Short examples showed the extendibility of the different subcomponents.
We presented a method to efficiently localize positions within a given mesh and described bindings of the library to scripting languages.
Finally the practical usability of the library was demonstrated by a source analysis of experimental data of a somatosensory stimulation.
The DUNEuro toolbox offers a flexible and efficient way to solve the EEG/MEG forward problem using modern mathematical methods.
There are several open goals regarding the software implementation.
Foremost, a direct comparison with existing tools for computing forward solutions for EEG and MEG, such as the SimBio toolbox, is currently performed.
Similar to the latter toolbox, a closer integration into existing EEG/MEG source analysis frameworks \cite{Oostenveld2011,Tadel2011,Gramfort2013} would further facilitate its usability.
An integration into the Brainstorm toolbox \cite{Tadel2011}, for instance, is currently under development (\url{https://neuroimage.usc.edu/brainstorm/Tutorials/Duneuro})\cite{Medani2021}.
This would then also allow to evaluate the advantages and disadvantages of the new FEM methods that are now available through the DUNEuro code in practical applications.
Additionally, this integration would offer the use of DUNEuro for different inverse approaches.
Several other forward problems, e.g., electric or magnetic brain stimulation, are already partly implemented, but their support should be improved and evaluated.
Of special interest would then be a connection to optimization procedures for transcranial direct current stimulation \cite{Dmochowski2011, Wagner2016, Khan2019}.
In order to improve the stability of the codebase and ensure the reliability of the results even under future modifications, a testing framework using continuous integration should be implemented.

\section*{Acknowledgments}
We thank Andreas Wollbrink for technical assistance and his valuable advice for the experiment and   Karin Wilken, Hildegard Deitermann and Ute Trompeter for their help with the EEG/MRI data collection.

A.W. and C.E. were supported by the DFG (https://www.dfg.de/en) through the Cluster of Excellence 1003 (EXC 1003 Cells in Motion). S.S. and C.H.W. were supported by the DFG through project WO1425/7-1 and C.H.W. additionally by the DFG priority program SPP1665, project WO1425/5-2. M.C.P. and C.H.W. were supported by the EU project ChildBrain (Marie Curie Innovative Training Networks, grant agreement 641652, http://www.childbrain.eu). S.P. and T.M. were supported by the Academy of Finland key project (305055), T.M. by the tenure track funding by Tampere University, and S.P. by the AoF Centre of Excellence of Inverse Modelling
and Imaging (334465), see https://www.aka.fi/en/. S.S., S.P, and C.H.W. were supported by a bilateral (Münster-Tampere) researcher exhange project by DAAD (57405052) and AoF (317165) and S.P. and C.H.W. by its successor by DAAD (57523877) and AoF (334465), see https://www.daad.de/en and https://www.aka.fi/en. J.V. was supported by the Austrian Wissenschaftsfonds (FWF), project I 3790-B27. We would also like to gratefully mention Germany's Excellence Strategy (EXC 2044-390685587), Mathematics Münster: Dynamics-Geometry-Structure.

The funders had no influence on study design, data collection and analysis, decision to publish, or preparation of the manuscript.

\ifarxiv
\else
\nolinenumbers
\fi

\bibliography{literature}

% mark the end of the document, to get a page number of the last page, before the delayed floats
\label{page:end}

\end{document}